\documentclass[11pt]{article}	% The documentclass must be ``report''.
\usepackage{amsmath,amsthm,amsfonts,amscd, amsmath}	
\usepackage{graphicx}
\usepackage{setspace}
\usepackage{color}
\usepackage{enumerate}
\usepackage{ upgreek }
\usepackage{authblk}
\usepackage{url}		% Allows good typesetting of web URLs.
\usepackage{tcolorbox}
\usepackage{multirow}

\usepackage{fancyhdr}
\begin{document}
%========================================================================
% First Page
%========================================================================
\title{
Technical Note:
Continuum Theory of Mixture for Three-phase Thermomechanical Model of Fiber-reinforced Aerogel Composites
}

\author{Pratyush Kumar Singh and Danial Faghihi\\
Department of Mechanical and Aerospace Engineering, University at Buffalo}

\maketitle

%\tableofcontents

%========================================================================
% Document body
%========================================================================
%\clearpage

%++++++++++++++++++++++++++++++++++++++++++++++++++++++++++++++++++++++++
\section{Introduction}
We present a thermodynamically consistent three-phase model for the coupled thermal transport and mechanical deformation of ceramic aerogel porous composite materials, which is formulated via continuum mixture theory. The composite comprises a solid silica skeleton, a gaseous fluid phase, and dispersed solid fibers. The thermal transport model incorporates the effects of meso- and macro-pore size variations due to the Knudsen effect, achieved by upscaling phonon transport relations to derive constitutive equations for the fluid thermal conductivity. The mechanical model captures solid-solid and solid-fluid interactions through momentum exchange between phases. A mixed finite element formulation is employed to solve the multiphase model, and numerical studies are conducted to analyze key features of the computational model.

%++++++++++++++++++++++++++++++++++++++++++++++++++++++++++++++++++++++++
\section{Continuum Theory of Mixture}
In this section, we present a theoretical framework based on the continuum theory of mixtures \cite{BOWEN19801129,ERINGEN1965197} to model the thermo-mechanical behavior of fiber-reinforced aerogel composite materials. The mixture theory is formulated for a continuum consisting of $M$ phases, each occupying the same physical space simultaneously, with the reference configuration being the state in which all phases are present.
The body experiences a deformation that transforms the reference configuration into a current configuration, where the position of material points is denoted by $\boldsymbol{x}$ at time $t$ defined via the motion $\mathcal{X}_{\alpha}$ as follows,
\begin{equation}
    \boldsymbol{x} = {\mathcal{X}}_{\alpha} (\boldsymbol{X}_{\alpha},t), \quad \alpha =1,2,...,M 
\end{equation}
where ${\boldsymbol{X}}_{\alpha}$ is the material point positions of the $\alpha$th phase. The volume fraction of the $\alpha$th phase is,
\begin{equation}
    \phi_{\alpha} (\boldsymbol{x}, t) = \frac{d V_{\alpha}}{d V} \quad \text{with} \quad \sum_{\alpha} \phi_{\alpha} = 1
\end{equation}
where $dV$ represents a differential volume element containing the point $\boldsymbol{x}$, and ${dV}_{\alpha}$ refers to the fraction of the volume occupied by phase $\alpha$. Moreover, the velocity $\mathbf{v}_{\alpha}$ and symmetric part of the velocity gradient $\mathbf{D}_{\alpha}$ of each phase are defined as follows
\begin{equation}
    \mathbf{v}_{\alpha}(\boldsymbol{x},t) = \frac{\partial \mathcal{X}_{\alpha} (\boldsymbol{X}_{\alpha},t)}{\partial t}, \quad \mathbf{D}_{\alpha} = \frac{1}{2} (\nabla \mathbf{v}_{\alpha} + \nabla {\mathbf{v}}_{\alpha}^{T} )
\end{equation}

The balance laws for each phase of the mixture are:

\noindent
\textit{1 - Balance of Mass}
\begin{equation}
\frac{\partial (\rho_\alpha \phi_\alpha)}{\partial t}  + \nabla \cdot (\rho_\alpha \phi_\alpha \mathbf{v}_\alpha) = S_\alpha.
\label{eq:mb}
\end{equation}
where $\rho_{\alpha}$ is the mass density of each phase $\alpha$. 

\noindent
\textit{2 - Balance of Linear Momentum}
\begin{equation}
\rho_\alpha \phi_\alpha \frac{d^\alpha}{dt} (\mathbf{v}_\alpha) = \nabla\cdot \mathbf{T}_\alpha + \rho_\alpha \phi_\alpha \mathbf{b}_\alpha + \mathbf{m}_\alpha,
\end{equation}
where $\mathbf{m}_\alpha$ is the momentum supplied by other phases to the $\alpha$th phase,
$\mathbf{M}_\alpha$ is the intrinsic moment of momentum vector for phase $\alpha$, 
and the material time-derivative of any differentiable function $\vartheta$ is,
\begin{equation}
    \frac{d^{\alpha} \vartheta}{dt} = \frac{\partial \vartheta }{\partial t} + \mathbf{v}_{\alpha} \cdot \nabla \vartheta 
\end{equation}

\noindent
\textit{3 - Balance of energy}
\begin{equation}
\rho_\alpha \phi_\alpha \frac{d^\alpha}{dt} (\varepsilon_\alpha) = 
\mathbf{T}_\alpha : \mathbf{D}_\alpha - \nabla \cdot \mathbf{q}_\alpha + \rho_\alpha \phi_\alpha {r}_\alpha +  {e}_\alpha.
\label{eq:energy_balance}
\end{equation}
Here, $\boldsymbol{T}_{\alpha}$ represents the partial Cauchy stress tensor,
$\boldsymbol{b}_{\alpha}$ denotes the body force per unit mass, $\boldsymbol{q}_{\alpha}$ is the partial heat flux, and $r_{\alpha}$ refers to the external heat supply, associated with the $\alpha$th phase.
 Furthermore, the interactions between the phases are defined by $S_{\alpha}$, $\mathbf{m}_{\alpha}$, and
$e_{\alpha}$ that are the mass, momentum, and energy supplied to the
phase $\alpha$ by other phases.\\

One can define the balance laws for the mixture for the mass density $\rho$, the stress $\mathbf{T}$, the body force per unit mass $\mathbf{b}$, the internal energy per unit mass $\varepsilon$, heat supply, and the heat flux $\mathbf{q}$ for the mixture 
as,
\begin{eqnarray}
\frac{\partial \rho}{\partial t} 	&+& \nabla \cdot (\rho \mathbf{v}) = 0\\  
\rho \frac{ d\mathbf{v}}{dt}  	&=& \nabla \cdot \mathbf{T} + \rho \mathbf{b}\\
\rho \frac{d}{dt} (\varepsilon) 	&= &
\mathbf{T} : \mathbf{D} - \nabla \cdot \mathbf{q} + \rho {r}.
\end{eqnarray}
where $\rho$ is the mass density of the mixture. If we sum the phase balance laws over all phases, the sums are compatible with the above relations if,

\begin{eqnarray}
\mathbf{T} &=& \sum_\alpha \left( \mathbf{T}_\alpha - \rho_\alpha\phi_\alpha \mathbf{p}_\alpha \otimes  \mathbf{p}_\alpha \right)\\
\mathbf{b} & = & \frac{1}{\rho} \sum_\alpha\rho_\alpha\phi_\alpha \mathbf{b}_\alpha\\
{r} & = & \frac{1}{\rho}\sum_\alpha \rho_\alpha\phi_\alpha {r}_\alpha\\
\mathbf{q} & = & \sum_\alpha \left(\mathbf{q}_\alpha - \mathbf{T}_\alpha^T \mathbf{p}_\alpha + 
 \rho_\alpha\phi_\alpha \varepsilon_\alpha \mathbf{p}_\alpha + \frac{1}{2} \rho_\alpha\phi_\alpha \mathbf{p}_\alpha (\mathbf{p}_\alpha \cdot \mathbf{p}_\alpha) \right)
\end{eqnarray}
where $\mathbf{p}_\alpha$ is the diffusion velocity, defined as
\begin{equation}
\mathbf{p}_\alpha = \mathbf{v}_\alpha - \mathbf{v}, \quad\qquad\qquad 
\sum_\alpha \rho_\alpha\phi_\alpha \mathbf{p}_\alpha = 0,
\end{equation}
and the mixture velocity $\mathbf{v}$ is the mass-averaged velocity,
\begin{equation}
\mathbf{v} = 
\frac{1}{\rho}  \rho_\alpha\phi_\alpha \mathbf{v}_\alpha.
\end{equation}
%
%--------------------------Mechanical Theory-----------------------------
%------------------------------------------------------------------------
\section{Mechanical Theory of Composite}
In this paper, we consider the fiber-aerogel composite consisting of three phases, where the index $s,g$, and $f$ represents the solid silica skeleton phase, gaseous fluid phase, and solid fiber phase, respectively.
The sum of volume fractions follows the relation,
\begin{equation}
    \phi_s + \phi_g + \phi_f = 1
    \label{eq:volume_fraction}
\end{equation}
Assuming the fluid has no average shear viscosity, the stress in fluid, solid aerogel, and fiber phases are obtained as,
\begin{eqnarray}
\mathbf{T}_g &=& -\phi_g p \mathbf{I}\\
\mathbf{T}_s &=& -\phi_s p \mathbf{I} + \mathbf{T}'_s\\
\mathbf{T}_f &=& -\phi_f p \mathbf{I} + \mathbf{T}'_f,
\end{eqnarray}
where $p$ is the equilibrium partial fluid pressure and $\mathbf{T}'_s$ is the effective solid stress and $\mathbf{T}'_f$ is the effective fiber stress.

Considering incompressible phases ($\rho_\alpha=$ constant) and
non-reactive mixture ($S_\alpha=0$),
the mass balance (\ref{eq:mb}) for solid silica and fiber can be written as,
\begin{eqnarray}
\frac{\partial \phi_s}{\partial t} + \phi_s \nabla \cdot \mathbf{v}_s = 0,
\quad
\frac{\partial \phi_f}{\partial t} + \phi_f \nabla \cdot \mathbf{v}_f = 0.
\label{eq:mb_solid_fibers}
\end{eqnarray}
Adding the equations in \eqref{eq:mb_solid_fibers}, and using \eqref{eq:volume_fraction}, we get,
\begin{equation}
\frac{\partial \phi_g}{\partial t} = \phi_s \nabla \cdot \mathbf{v}_s + \phi_f \nabla \cdot \mathbf{v}_f.
\label{eq:gen_eq_state}
\end{equation}

%------------------------Constitutive Relations--------------------------
\subsection{Constitutive Relations}

Assuming linear isotropic elasticity for the skeleton and fiber yields
\begin{equation}
\mathbf{T}'_s = 2\mu_s \boldsymbol{E}_s + \lambda_s \text{tr}(\boldsymbol{E}_s) \mathbf{I}, 
\quad
\mathbf{T}'_f = 2\mu_f \boldsymbol{E}_f + \lambda_f \text{tr}(\boldsymbol{E}_f) \mathbf{I},
\end{equation}
where $\boldsymbol{E}_s$ and $\boldsymbol{E}_f$ are the skeleton and fiber strains and computed from the corresponding deformations, $\mathbf{u}_s$ and $\mathbf{u}_f$, as,
\begin{equation}
\boldsymbol{E}_s  = \frac{1}{2} (\nabla \mathbf{u}_s + \nabla \mathbf{u}_s^T),
\quad
\boldsymbol{E}_f  = \frac{1}{2} (\nabla \mathbf{u}_f + \nabla \mathbf{u}_f^T).
\end{equation}
The momentum interactions between the fluid, aerogel skeleton, and fibers are assumed to be, 
\begin{equation}
\mathbf{m}_s = \gamma_s (\mathbf{v}_s - \mathbf{v}_g) + {\chi} \nabla \cdot( \boldsymbol{E}_s - \boldsymbol{E}_f)
\end{equation}
\begin{equation}
\mathbf{m}_g = \gamma_s (\mathbf{v}_g - \mathbf{v}_s) + \gamma_f (\mathbf{v}_g - \mathbf{v}_f)
\end{equation}
\begin{equation}
\mathbf{m}_f = {\chi} \nabla \cdot(\boldsymbol{E}_f - \boldsymbol{E}_s) + \gamma_f ( \mathbf{v}_f - \mathbf{v}_g)
\end{equation}
where $\gamma_s$ and $\gamma_f$ are the drag coefficients for the aerogel skeleton and the fibers, and ${\chi}$ is the coefficient for strain coupling between solid aerogel and fibers. The momentum interaction between skeleton and fibers is assumed to be dependent on the difference in strain, inspired by various studies related to fiber composites \cite{NAIRN199763, xia2002shear, beyerlein1996comparison}.
The interaction coefficient $\chi$ for strain coupling depends on the magnitude of relative strain between the fiber and the solid. When the magnitude of relative strain becomes above a certain threshold $\epsilon_{strain}$, the interaction coefficient $\chi$ can be considered zero as it is a case of debonding as there can be no more strain transfer between the fiber and solid. The interaction coefficient can be written as,
 \begin{equation}
\chi = 
\begin{cases}
\chi_0 \quad \text{if} \quad ||\boldsymbol{E}_s - \boldsymbol{E}_f||< \epsilon_{strain} \\
0 \: \: \quad \text{if} \quad ||\boldsymbol{E}_s - \boldsymbol{E}_f||> \epsilon_{strain}
\end{cases}
 \end{equation}
Considering the above assumptions, the linear momentum equation for solid skeleton and fiber, after ignoring the body force and inertia, becomes,
\begin{eqnarray}
\nabla \cdot \mathbf{T}'_s - \phi_s \nabla p + \gamma_s (\mathbf{v}_s - \mathbf{v}_g) + {\chi} \nabla \cdot(\boldsymbol{E}_s - \boldsymbol{E}_f) &=& 0,\\
\nabla \cdot \mathbf{T}'_f - \phi_f \nabla p + \gamma_f (\mathbf{v}_f - \mathbf{v}_g) + {\chi} \nabla \cdot(\boldsymbol{E}_f - \boldsymbol{E}_s) &=& 0.
\end{eqnarray}
The linear momentum equation for fluid, after ignoring the body force and inertia, is
\begin{equation}
- \phi_g \nabla p + \gamma_s (\mathbf{v}_g - \mathbf{v}_s) + \gamma_f(\mathbf{v}_g - \mathbf{v}_f) = 0,
\end{equation}
and reduce to Darcy's law:
\begin{equation}
\boxed{
\mathbf{G} = \phi_g \mathbf{v}_g = - k \nabla p,
}
\label{eq:Darcy}
\end{equation}
where $k$ is the permeability depending on the aerogel skeleton and fiber permeability values and their volume fractions \cite{polym15030728, NAIK201422}.
Finally, the constitutive relation for gaseous fluid density $\rho_f(p)$, under isothermal and compressible fluid assumption, can be written as
\begin{equation}\label{eq:fluid_density}
\frac{\partial \rho_g}{\partial t} = 
C_0\frac{\rho_g}{\phi_g} \frac{\partial p}{\partial t},
\end{equation}
where $C_0$ is similar to storage coefficient\cite{phillips2005finite}.
\subsection{Governing equations}
The mass balance for compressible fluid can be written as,
\begin{eqnarray}
    \frac{\partial (\rho_g \phi_g) }{\partial t} + \nabla . (\rho_g \phi_g \mathbf{v}_g) = 0\\
     \phi_g \frac{\partial \rho_g}{\partial t} + \rho_g \frac{\partial \phi_g}{\partial t} + \rho_g \nabla \cdot ( \phi_g \mathbf{v}_g ) = 0
     \label{eq:fluid_mb}
\end{eqnarray}
Using the relation \eqref{eq:gen_eq_state} in \eqref{eq:fluid_mb} and the constitutive relation \eqref{eq:fluid_density} we get,
\begin{eqnarray}
    C_0 \phi_g \frac{\rho_g}{\phi_g} \frac{\partial p}{\partial t} + \rho_g \phi_s \nabla \cdot \mathbf{v}_s + \rho_g \phi_f \nabla \cdot \mathbf{v}_f + \rho_g \nabla \cdot (\phi_g \mathbf{v}_g) = 0\\
     C_0 \frac{\partial p}{\partial t} + \phi_s \nabla \cdot \mathbf{v}_s + \phi_f \nabla \cdot \mathbf{v}_f + \nabla \cdot (\phi_g \mathbf{v}_g) =0
\end{eqnarray}
Finally, using \eqref{eq:Darcy} in the above equation, we get, 
\begin{eqnarray}
    C_0 \frac{\partial p}{\partial t} + \phi_s \frac{\partial}{\partial t} (\nabla \cdot \mathbf{u}_s) + \phi_f \frac{\partial}{\partial t} (\nabla \cdot \mathbf{u}_f) - \nabla \cdot ( k \nabla p) = 0 
\end{eqnarray}
%----------------------------Thermal Theory------------------------------
%------------------------------------------------------------------------
\section{Thermal Transport Theory of Composite}

Defining the specific heat capacity at constant volume as $c_\alpha = \frac{\partial \varepsilon_\alpha}{\partial \theta_\alpha}$ and replacing the material derivative in (\ref{eq:energy_balance}) with partial derivative, we get
\begin{equation}
\rho_\alpha \phi_\alpha c_\alpha \left[ \frac{\partial \theta_\alpha}{\partial t} + (\mathbf{v}_\alpha \cdot \nabla \theta_\alpha) \right]= 
\mathbf{T}_\alpha : \mathbf{D}_\alpha - \nabla \cdot \mathbf{q}_\alpha + \rho_\alpha \phi_\alpha {r}_\alpha +  {e}_\alpha, \quad \alpha = s, f, g.
\end{equation}

\subsection{Constitutive Relations}

Assuming the Fourier law of heat conduction for the solid and fluid partial heat flux, we have
\begin{eqnarray}
\mathbf{q}_s &=& - \phi_s \kappa_s \nabla \theta_s,\\
\mathbf{q}_g &=& - \phi_g \kappa_g \nabla \theta_g.\\
\mathbf{q}_f &=& - \phi_f \kappa_f \nabla \theta_f
\end{eqnarray}
The internal energy supplies ${e}_{\alpha}$ account for the heat exchanges between the phases and their relative movements. Nonlinear thermal contact resistance is incorporated to capture microscale effects in solid-fluid heat exchange, where the heat transfer rate varies depending on the temperature difference between the phases \cite{AMIRI1994939, YANG20104316}. For our model, we consider the below constitutive relations,
\begin{eqnarray}
{e}_s = h_{sg} (\theta_s - \theta_g)^3 + h_{sf} (\theta_s - \theta_f)^3\\
{e}_g = h_{sg} (\theta_g - \theta_s)^3 + h_{gf} (\theta_g - \theta_f)^3\\
{e}_f = h_{sf} (\theta_f - \theta_s)^3 + h_{gf} (\theta_f - \theta_g)^3\\
\end{eqnarray}
where $h_{sg}, h_{sf},h_{gf} >0$ in general depends on the thermal properties and velocity fields to account for the convective heat transfer. The thermal conductivity of the fluid is characterized by the pore size $\omega$, which is modeled as, 
\begin{equation}
    \kappa_g = \frac{\kappa_{bg}}{\beta (l_g/\omega) +1}
    \label{eq:thermal conductivity}
\end{equation}
where $\kappa_{bg}$ is the conductivity of gas in free space, $l_g$ is a length scale, and $\beta$ is a non-dimensional parameter. The fluid thermal conductivity defined using the expression \eqref{eq:thermal conductivity} is adopted from kinetic theory to capture the effect of Knudsen effect due to phonon transport \cite{majumdar1993}. The fundamental equations governing heat transfer can be expressed by incorporating the constitutive relations and disregarding the external heat source. 
\begin{equation}
\begin{aligned}
&\rho_s \phi_s c_s \left[ \frac{\partial \theta_s}{\partial t} + (\mathbf{v}_s \cdot \nabla \theta_s) \right]= 
\mathbf{T}_s : \mathbf{D}_s + \nabla \cdot (\phi_s \kappa_s \nabla \theta_s) +h_{sg} (\theta_s - \theta_g)^3 + h_{sf} (\theta_s - \theta_f)^3\\
&\rho_g \phi_g c_g \left[ \frac{\partial \theta_g}{\partial t} + (\mathbf{v}_g \cdot \nabla \theta_g) \right]= 
- \phi_g \:p \: \text{tr}(\mathbf{D}_g) + \nabla \cdot (\phi_g \kappa_g \nabla \theta_g) + h_{sg} (\theta_g - \theta_s)^3 + h_{gf} (\theta_g - \theta_f)^3\\
&\rho_f \phi_f c_f \left[ \frac{\partial \theta_f}{\partial t} + (\mathbf{v}_f \cdot \nabla \theta_f) \right]= 
\mathbf{T}_f : \mathbf{D}_f + \nabla \cdot (\phi_f \kappa_f \nabla \theta_f) +h_{sf} (\theta_f - \theta_s)^3 + h_{gf} (\theta_f - \theta_g)^3\\
\end{aligned}
\end{equation}

%======================
\section{Numerical Results}
We use continuous Galerkin finite element methods to solve the equations for thermomechanical models derived in the previous section.
The governing equations and their associated boundary conditions for the three-phase mechanical model derived in the previous section can be summarized as,
\begin{equation}
\begin{aligned}
     C \frac{\partial p}{\partial t} + \frac{\partial}{\partial t} (\nabla \cdot \boldsymbol{u}_s) + \frac{\partial}{\partial t} (\nabla \cdot \boldsymbol{u}_f) + \nabla \cdot \mathbf{G} = 0 \quad in \quad \Omega^M &\\
    \mathbf{G} = -k \nabla p \quad in \quad \Omega^M &\\
     \nabla \cdot \boldsymbol{T}_s (\boldsymbol{u}_s) + \chi \nabla \cdot (\boldsymbol{E}_s - \boldsymbol{E}_f) = 0 \quad in \quad \Omega^M &\\
     \nabla \cdot \boldsymbol{T}_f (\boldsymbol{u}_f) + \chi \nabla \cdot(\boldsymbol{E}_f - \boldsymbol{E}_s) = 0 \quad in \quad \Omega^M &\\
    \boldsymbol{u}_s = \Bar{\boldsymbol{u}}_s \quad on \quad \Gamma_t^M &\\
    \boldsymbol{u}_s = \boldsymbol{0} \quad on \quad \Gamma_b^M &\\
    p = 0 \quad on \quad \Gamma^M &\\
    \mathbf{G} = \boldsymbol{0} \quad on \quad \Gamma^M &.
\end{aligned}
\label{eq:mech_pdes}
\end{equation}
The variational formulation for finite element solution of \eqref{eq:mech_pdes} can be written as,
\begin{equation}
\begin{aligned}
    &\text{Find} \: (p,\mathbf{G},\mathbf{u}_s,\mathbf{u}_f) \: \in \: \mathcal{W}_p \times \mathcal{W}_G \times \mathcal{W}_{u_s} \times \mathcal{W}_{u_f}, \text{such that}\\
    &\int_{\Omega^M} ( C p^{n+1} + \nabla \cdot \mathbf{u}_{s}^{n+1} + \nabla \cdot \mathbf{u}_{f}^{n+1}) w_p d \, \boldsymbol{x} - \int_{\Omega^M} \frac{\Delta t}{k} \mathbf{G}^{n+1} \cdot \boldsymbol{w}_G d \,\boldsymbol{x} + \int_{\Omega^M} \Delta t\: p^{n+1} \: \nabla \cdot \boldsymbol{w}_G d \, \boldsymbol{x} = 0\\
    &\int_{\Omega^M} \mathbf{T}_s(\mathbf{u}_{s}^{n+1}) : \nabla \boldsymbol{w}_s d\, \boldsymbol{x} 
    - \int_{\Omega^M} {\chi}(\mathbf{E}_s - \mathbf{E}_f) : \nabla \boldsymbol{w}_s d\, \boldsymbol{x} = 0\\
    &\int_{\Omega^M} \mathbf{T}_f(\mathbf{u}_{f}^{n+1}) : \nabla \boldsymbol{w}_f d\, \boldsymbol{x} - \int_{\Omega^M} {\chi} (\mathbf{E}_f - \mathbf{E}_s) : \nabla \boldsymbol{w}_f d\, \boldsymbol{x}= 0,
\end{aligned}
\end{equation}
for all test functions $w_p, \boldsymbol{w}_G, \boldsymbol{w}_s ,\boldsymbol{w}_f$, where $\Delta t$ is the time step.

The governing equations defining the thermal transport model and the corresponding boundary conditions are
\begin{equation}
\begin{aligned}
     \rho_s \phi_s c_s \frac{\partial \theta_s}{\partial t} = \nabla \cdot (\phi_s \kappa_s \nabla \theta_s) + h_{sg}{(\theta_s - \theta_g)}^3  + h_{sf}{(\theta_s - \theta_f)}^3  \quad in \quad \Omega^T &\\
     \rho_g \phi_g c_g \frac{\partial \theta_g}{\partial t} = \nabla \cdot (\phi_g \kappa_g \nabla \theta_g) + h_{sg}{(\theta_g - \theta_s)}^3  + h_{fg}{(\theta_g - \theta_f)}^3  \quad in \quad \Omega^T &\\
     \rho_f \phi_f c_f \frac{\partial \theta_f}{\partial t} = \nabla \cdot (\phi_f \kappa_f \nabla \theta_f) + h_{fg}{(\theta_f - \theta_g)}^3  + h_{sf}{(\theta_f - \theta_s)}^3  \quad in \quad \Omega^T &\\
     - \phi_\alpha \kappa_\alpha \nabla \theta_\alpha = h_{air} (\theta_\alpha - \theta_{cold}), \;\;\;\; \alpha \in (s,g,f) \quad on \quad \Gamma_b^T &\\ 
     - \phi_\alpha \kappa_\alpha \nabla \theta_\alpha = h_{air} (\theta_\alpha - \theta_{hot}), \;\;\;\; \alpha \in (s,g,f) \quad on \quad \Gamma_c^T. &\\ 
\end{aligned}
\end{equation}
The corresponding variational formulation is written as,
\begin{equation}
\begin{aligned}
    &\text{Find} \: (\theta_s, \theta_g, \theta_f) \: \in \: \mathcal{Z} \times \mathcal{Z} \times \mathcal{Z}, \text{such that}\\
    & \int_{\Omega^T} \rho_s \phi_s c_s \theta_{s}^{n+1} z_s d \,\boldsymbol{x} + 
    \int_{\Omega^T} \Delta t \:\phi_s \kappa_s (\nabla \theta_{s}^{n+1} \cdot \nabla z_s) d \,\boldsymbol{x}
    - \int_{\Omega^T} \Delta t \: h_{sf} {(\theta_{f}^{n+1} - \theta_{s}^{n+1})}^3 z_s d \,\boldsymbol{x}\\
    &- \int_{\Omega^T} \Delta t \: h_{sg} {(\theta_{g}^{n+1} - \theta_{s}^{n+1})}^3 z_s d \,\boldsymbol{x} + \int_{\Gamma^T} \Delta t \phi_s h_{air} (\theta_{s}^{n+1} - \theta_{cold}) z_s d \, \boldsymbol{s} = \int_{\Omega^T} \rho_s \phi_s c_s \theta_{s}^{n} z_s d\, \boldsymbol{x}\\
    & \int_{\Omega^T} \rho_g \phi_g c_g \theta_{g}^{n+1} z_g d \,\boldsymbol{x} + 
    \int_{\Omega^T} \Delta t \:\phi_g \kappa_g (\nabla \theta_{g}^{n+1} \cdot \nabla z_g) d \,\boldsymbol{x}
    - \int_{\Omega^T} \Delta t \: h_{fg} {(\theta_{f}^{n+1} - \theta_{g}^{n+1})}^3 z_g d \,\boldsymbol{x}\\
    &- \int_{\Omega^T} \Delta t \: h_{fg} {(\theta_{f}^{n+1} - \theta_{g}^{n+1})}^3 z_g d \,\boldsymbol{x} + \int_{\Gamma^T} \Delta t \phi_g h_{air} (\theta_{g}^{n+1} - \theta_{cold}) z_g d \, \boldsymbol{s} = \int_{\Omega^T} \rho_g \phi_g c_g \theta_{s}^{n} z_g d\, \boldsymbol{x}\\
    & \int_{\Omega^T} \rho_f \phi_f c_f \theta_{f}^{n+1} z_f d \,\boldsymbol{x} + 
    \int_{\Omega^T} \Delta t \:\phi_f \kappa_f (\nabla \theta_{f}^{n+1} \cdot \nabla z_f) d \,\boldsymbol{x}
    - \int_{\Omega^T} \Delta t \: h_{fg} {(\theta_{g}^{n+1} - \theta_{f}^{n+1})}^3 z_f d \,\boldsymbol{x}\\
    &- \int_{\Omega^T} \Delta t \: h_{sf} {(\theta_{s}^{n+1} - \theta_{f}^{n+1})}^3 z_f d \,\boldsymbol{x} + \int_{\Gamma^T} \Delta t \phi_f h_{air} (\theta_{f}^{n+1} - \theta_{cold}) z_f d \, \boldsymbol{s} = \int_{\Omega^T} \rho_f \phi_f c_f \theta_{f}^{n} z_f d\, \boldsymbol{x}
\end{aligned}
\end{equation}
for all test functions $z_s, z_g ,z_f$.

\begin{figure}[ht]
\centering
    \includegraphics[width=0.5\linewidth]{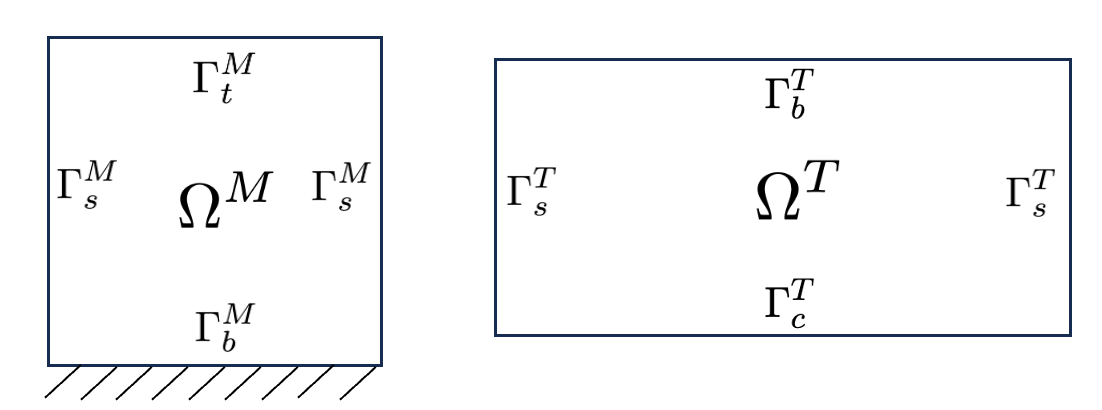}
    \vspace{-0.2in}
    \caption{Domains for (left) mechanical and (right) thermal transport models.}
    \vspace{-0.15in}
    \label{fig:domain_mechanical}
\end{figure}
A set of numerical experiments is conducted by implementing the above variational formulation in the finite element code FEniCS \cite{alnaes2015fenics}.
The domains and boundary conditions corresponding to the mechanical and thermal transport experiments are shown in Figure \ref{fig:domain_mechanical}. 

\begin{figure}[ht]
    \centering
    \begin{tabular}{c c c}
    \multicolumn{3}{c}{\includegraphics[width=0.3\linewidth]{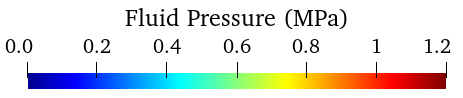}}\\
    \includegraphics[trim={3.5in 1.5in 3.5in 1.5in},clip,width=0.15\linewidth]{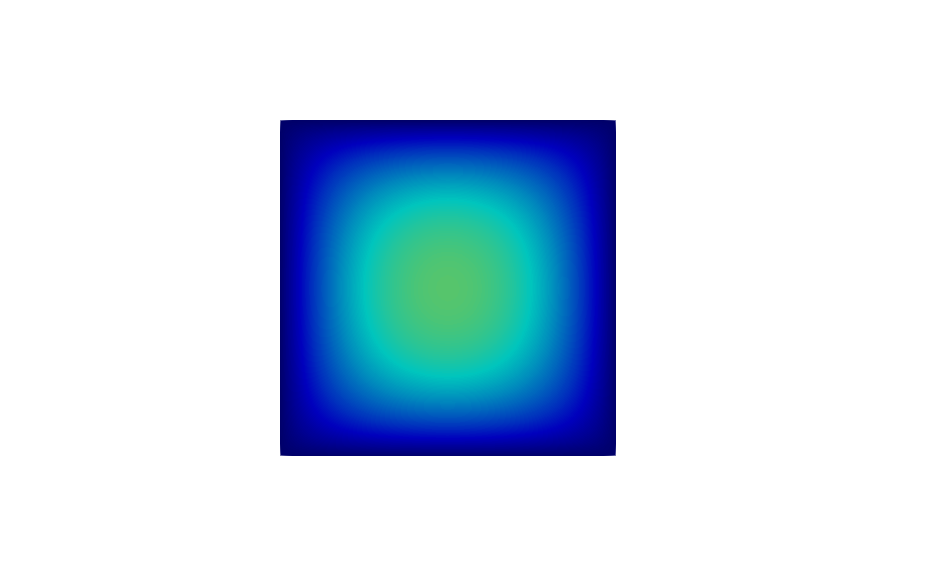}&
    \includegraphics[trim={3.5in 1.5in 3.5in 1.5in},clip,width=0.15\linewidth]{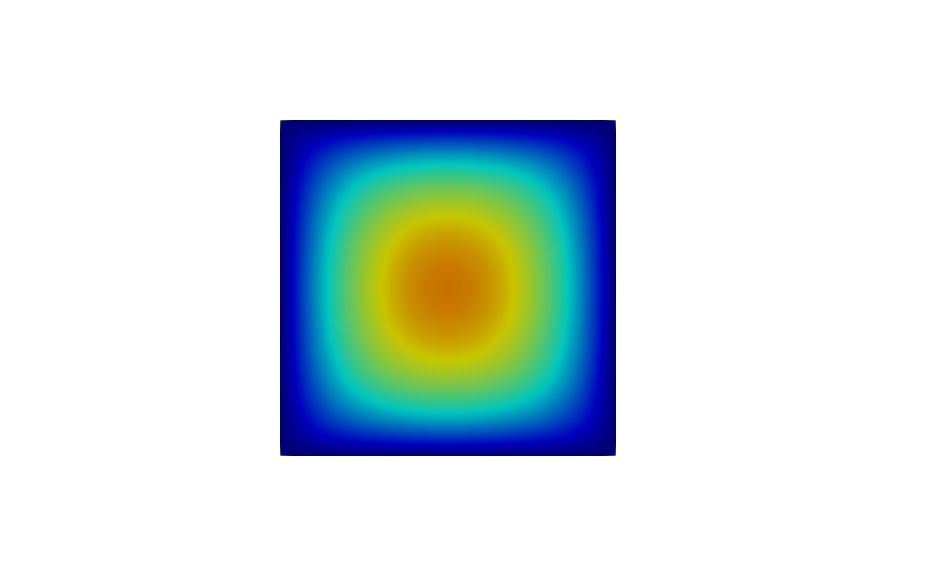}&
    \includegraphics[trim={3.5in 1.5in 3.5in 1.5in},clip,width=0.15\linewidth]{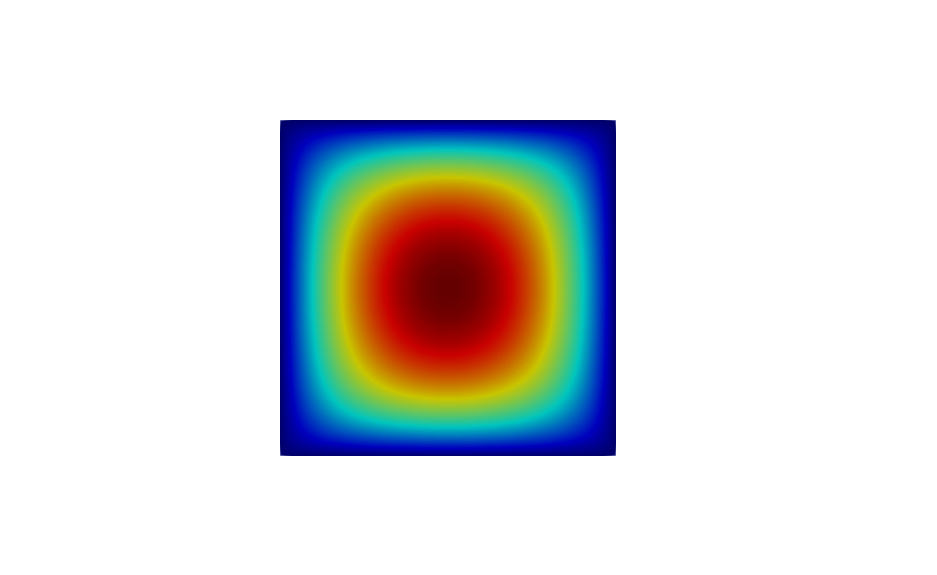}\\
    \multicolumn{3}{c}{\includegraphics[width=0.3\linewidth]{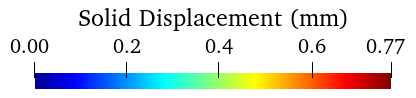}}\\
    \includegraphics[trim={3.5in 1.5in 3.5in 1.5in},clip,width=0.15\linewidth]{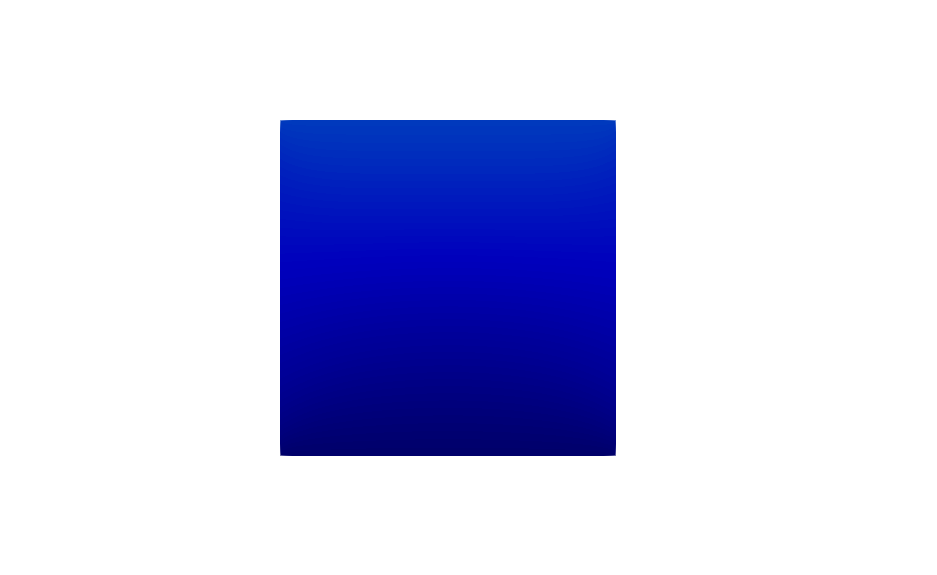}&
    \includegraphics[trim={3.5in 1.5in 3.5in 1.5in},clip,width=0.15\linewidth]{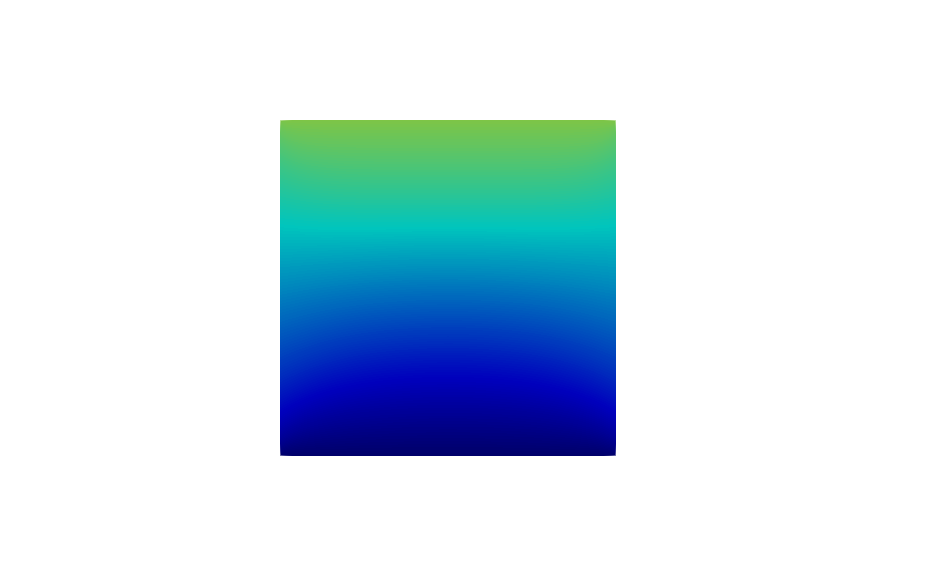}&
    \includegraphics[trim={3.5in 1.5in 3.5in 1.5in},clip,width=0.15\linewidth]{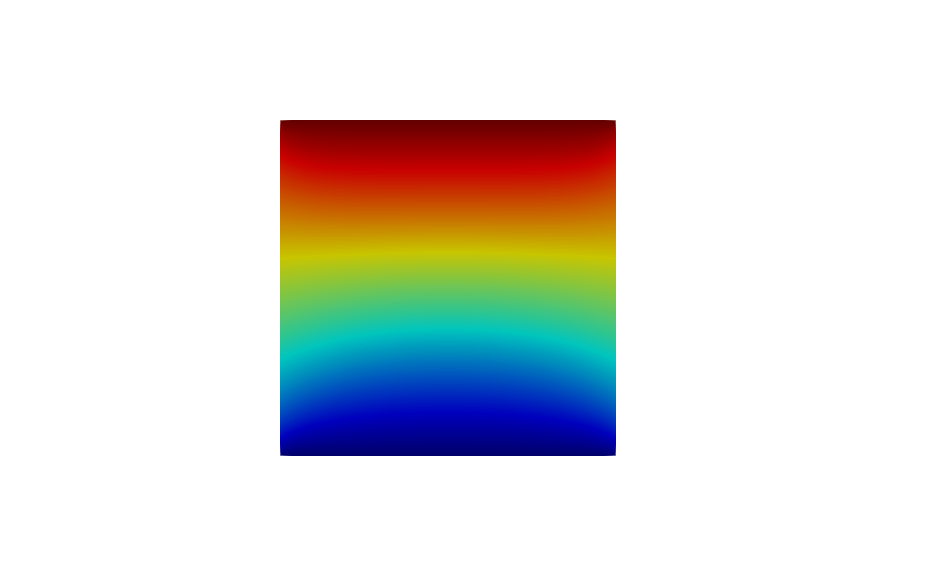}\\
    \multicolumn{3}{c}{\includegraphics[width=0.3\linewidth]{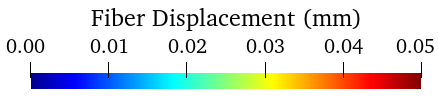}}\\
    \includegraphics[trim={3.5in 1.5in 3.5in 1.5in},clip,width=0.15\linewidth]{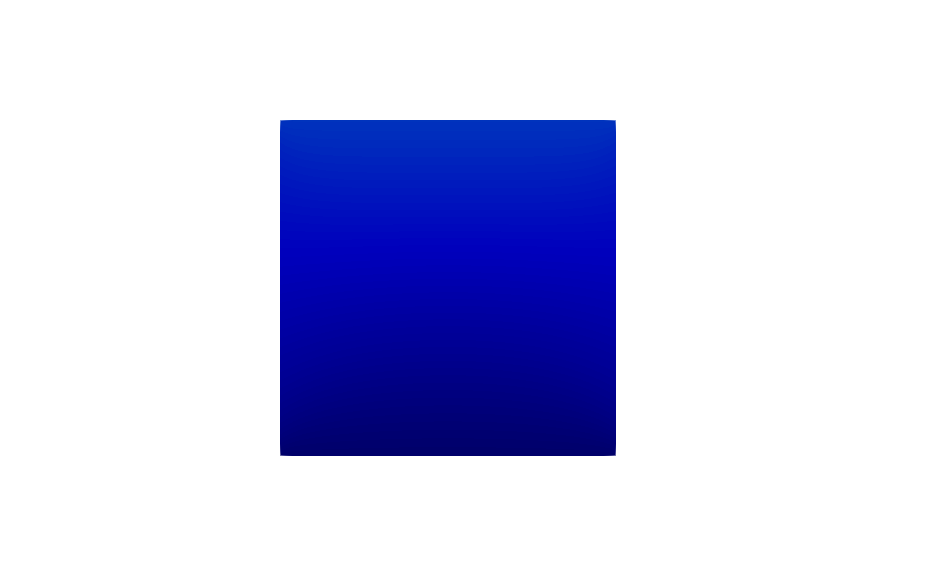}&
    \includegraphics[trim={3.5in 1.5in 3.5in 1.5in},clip,width=0.15\linewidth]{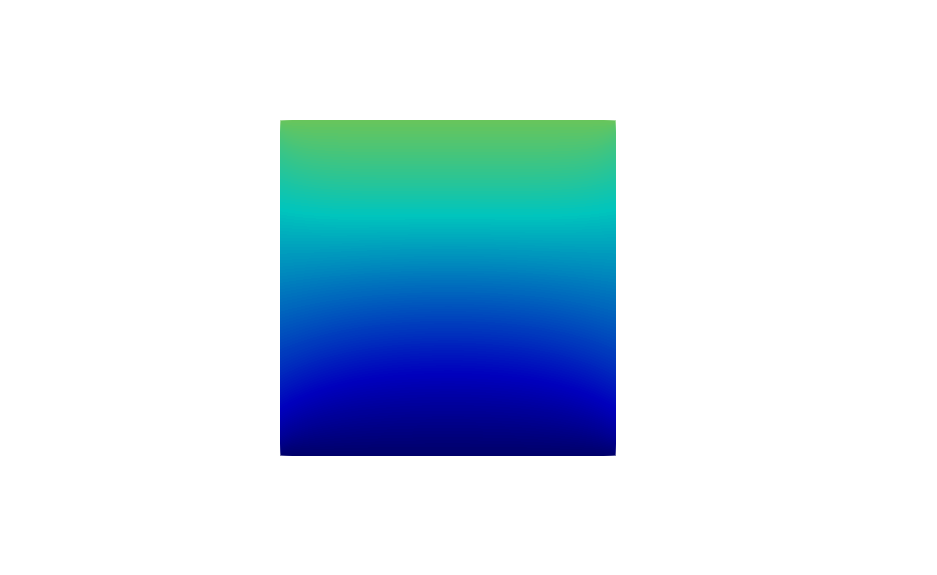}&
    \includegraphics[trim={3.5in 1.5in 3.5in 1.5in},clip,width=0.15\linewidth]{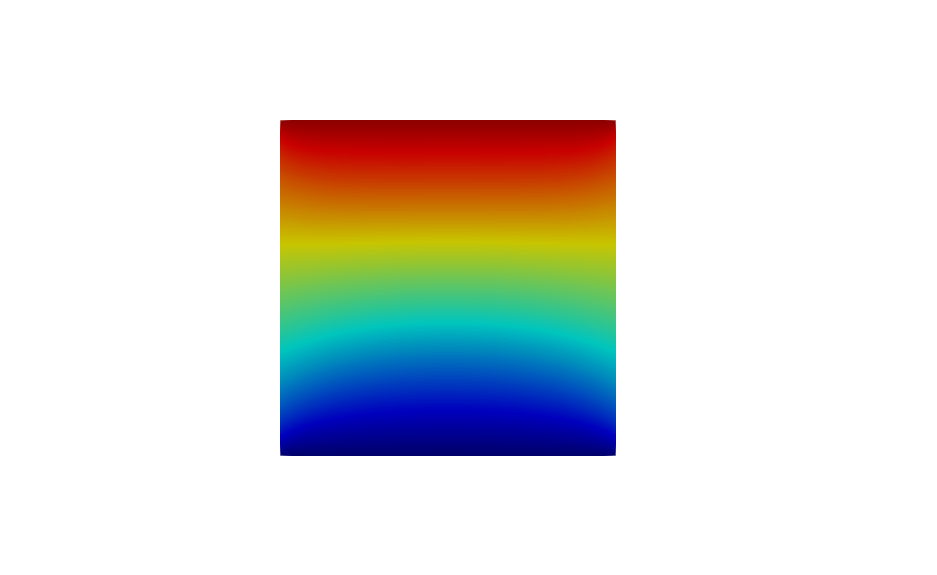}
    \end{tabular}
    \vspace{-0.1in}
    \caption{Contour plots of fluid pressure(top row), aerogel skeleton displacement (middle row) and fiber displacement (bottom row) at $t=$33 s, 66 s and 100 s from left to right.}
    \label{fig:deformation_plots}
\end{figure}

The numerical results of the composite deformation are shown in Figure \ref{fig:deformation_plots}, representing the time evolution of fluid pressure, solid
deformation and fiber deformation.
The length of the square side is $L= 1$ mm. The model parameters are $C = 8.5 \times 10^{-9}$ (1/Pa), $k = 1 \times 10^{-13} m^2$, $\lambda_s = 0.7 \: \text{MPa}$, $\mu_s = 0.27 \: \text{MPa}$, $\lambda_f = 5.77 \: \text{MPa}$ and $\mu_f = 3.84 \: \text{MPa}$. The interaction coefficient $\chi_{0} = 0.1 \: \text{MPa}$ for this scenario.

\begin{figure}[ht]
    \centering
    \begin{tabular}{c c c}
    \multicolumn{3}{c}{\includegraphics[width=0.3\linewidth]{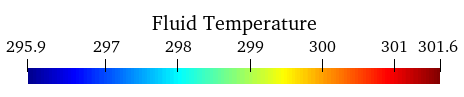}}\\
    \includegraphics[trim={1.9in 0.75in 1.9in 1.08in},clip,width=0.215\linewidth]{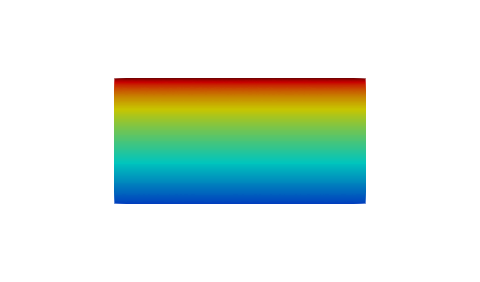}&
    \includegraphics[trim={3.5in 1.5in 3.5in 1.5in},clip,width=0.22\linewidth]{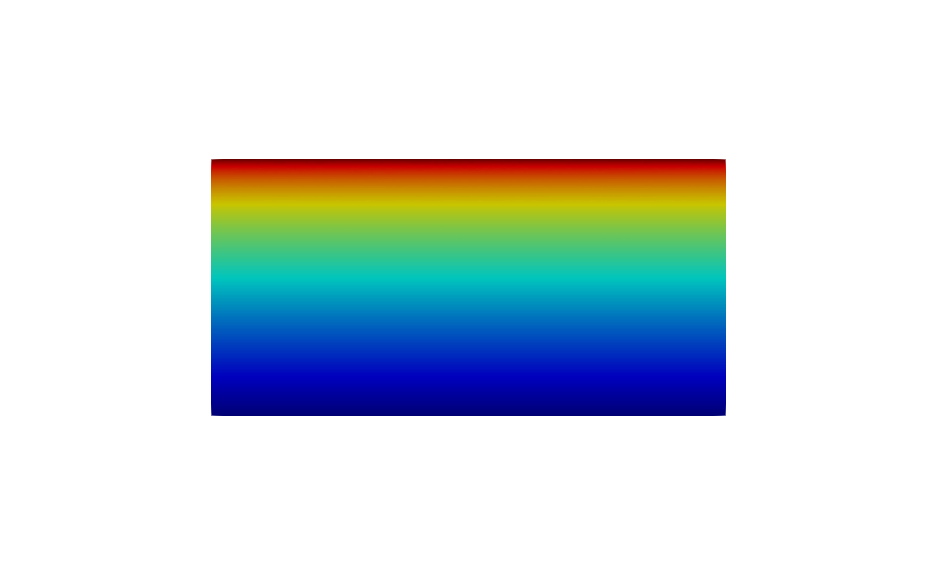}&
    \includegraphics[trim={3.5in 1.5in 3.5in 1.5in},clip,width=0.22\linewidth]{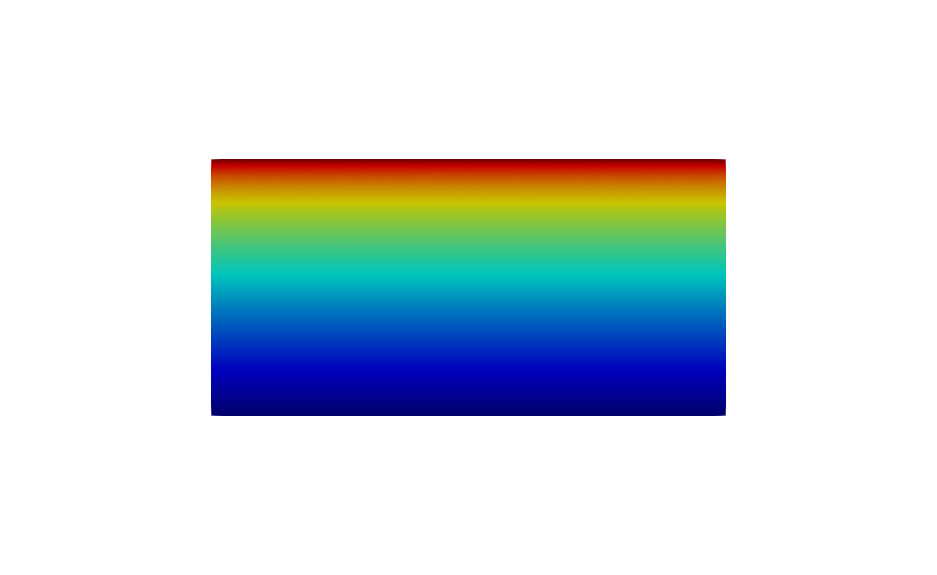}\\
    \multicolumn{3}{c}{\includegraphics[width=0.3\linewidth]{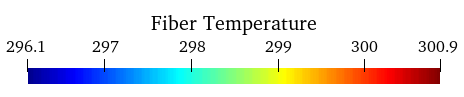}}\\
    \includegraphics[trim={3.5in 1.5in 3.5in 1.5in},clip,width=0.22\linewidth]{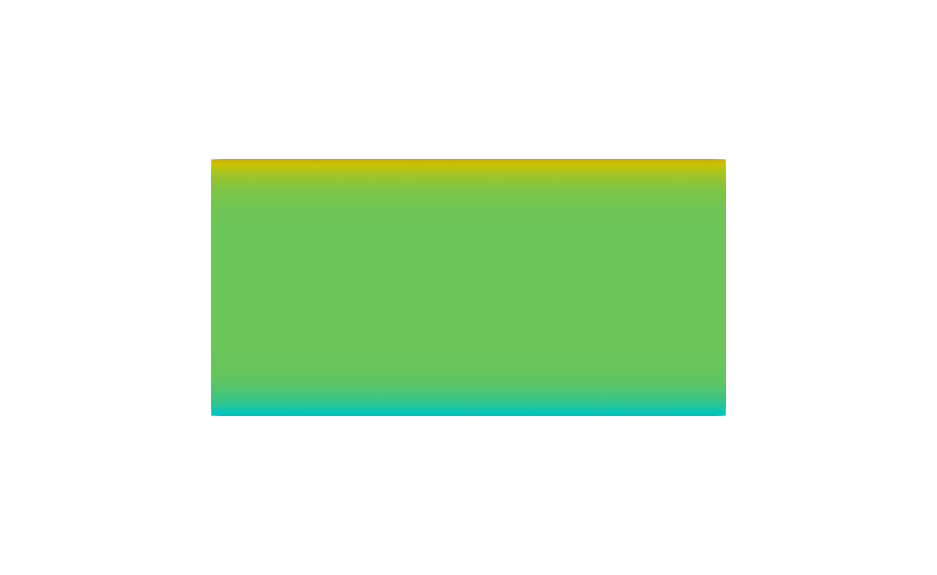}&
    \includegraphics[trim={3.5in 1.5in 3.5in 1.5in},clip,width=0.22\linewidth]{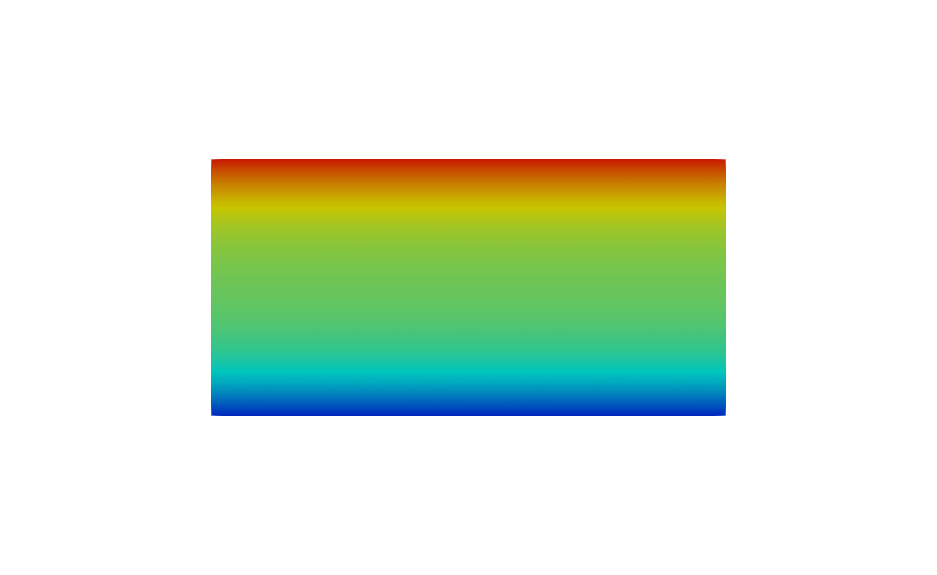}&
    \includegraphics[trim={3.5in 1.5in 3.5in 1.5in},clip,width=0.22\linewidth]{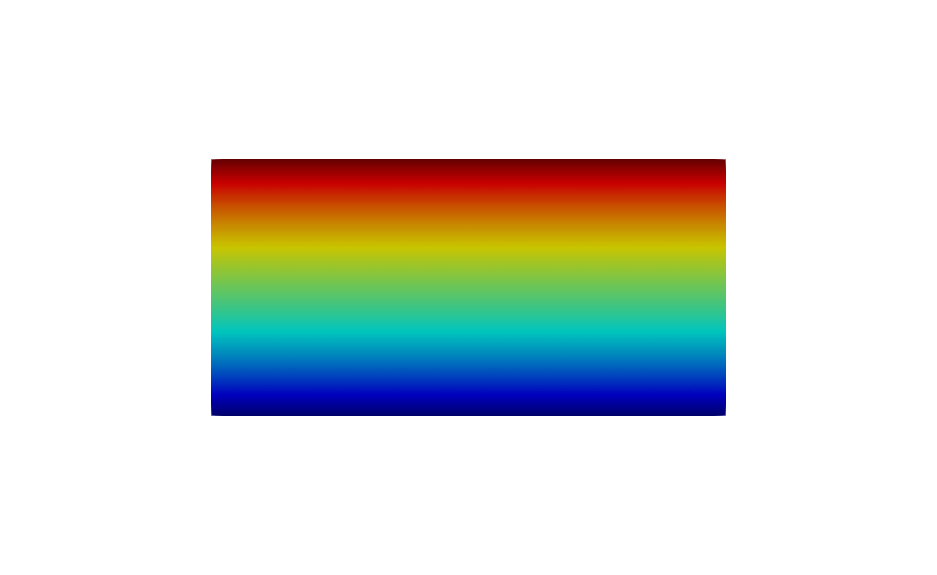}\\
    \multicolumn{3}{c}{\includegraphics[width=0.3\linewidth]{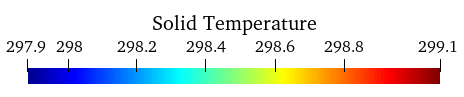}}\\
    \includegraphics[trim={3.5in 1.5in 3.5in 1.5in},clip,width=0.22\linewidth]{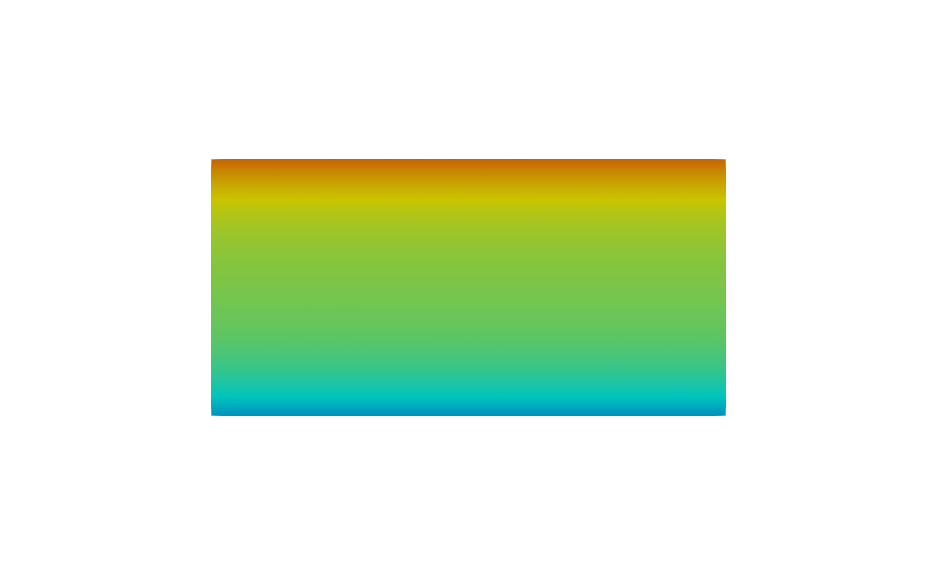}&
    \includegraphics[trim={3.5in 1.5in 3.5in 1.5in},clip,width=0.22\linewidth]{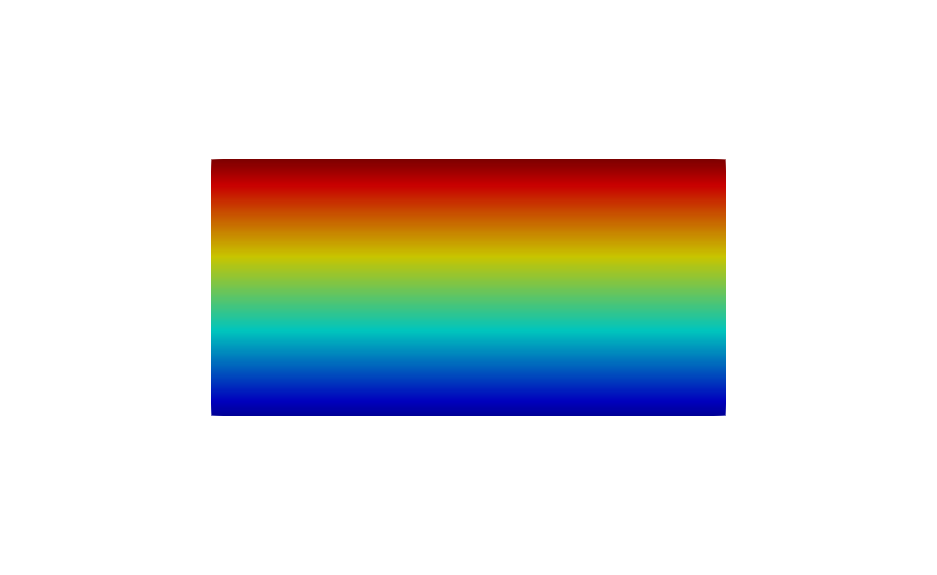}&
    \includegraphics[trim={3.5in 1.5in 3.5in 1.5in},clip,width=0.22\linewidth]{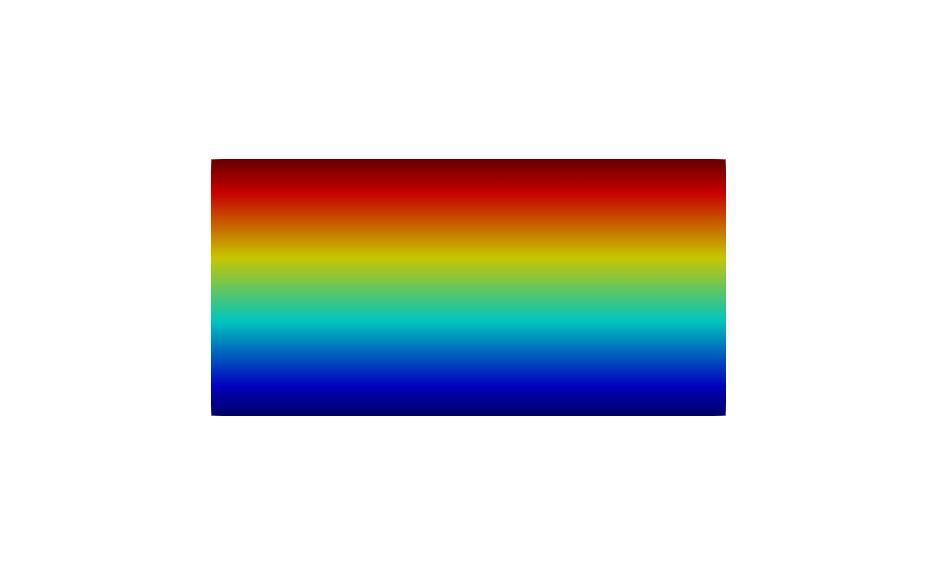}\\
    \end{tabular}
    \vspace{-0.1in}
    \caption{{Contour plots of the fluid temperature (top row), fiber temperature (middle row) and solid aerogel skeleton temperature (bottom row) at $t=$5s, 30s and 80s respectively from left to right.}}
    \label{fig:temperature_plots}
\end{figure}

\begin{figure}[ht]
    \centering
    \includegraphics[trim={0.5in 0.75in 0.8in 0.75in},clip,width=0.9\linewidth]{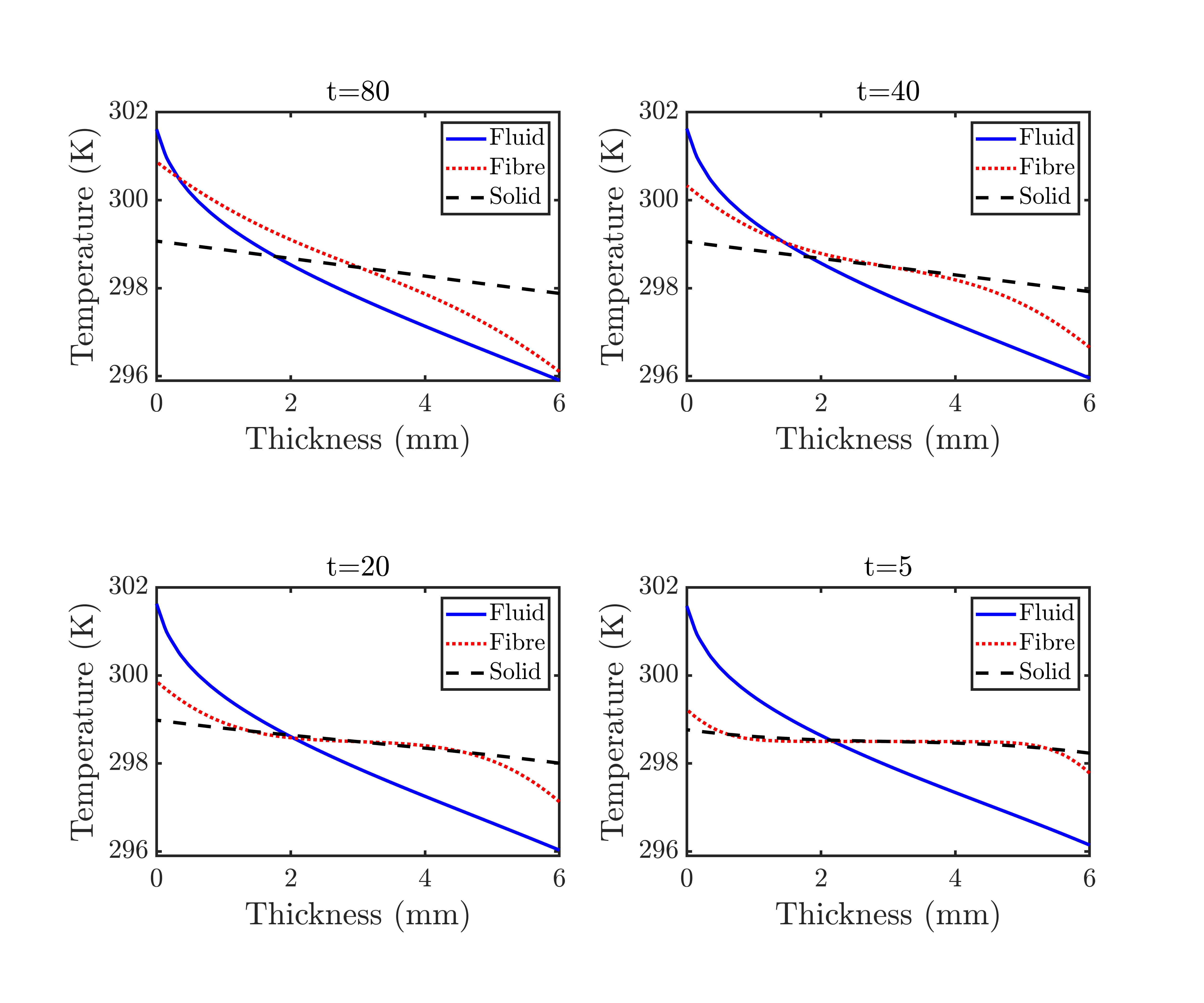}\\ 
    \caption{Temperature of all three phases along the thickness of domain through middle ($x = 0.5L$) across different times for $t=$ 5, 20, 40 and 80 s through numerical experiments for the heat transfer model}
    \label{fig:temperature_2}
\end{figure}

The results of the thermal transport model are shown in Figures \ref{fig:temperature_plots} and \ref{fig:temperature_2}, indicating the evolution of temperature through time for each of the three phases. 
The length of the domain is taken as $L = 12$ mm, and the thickness is $th = 6 \:$ mm. The thermal conductivities are $\kappa_s = 0.5 \:{(W/mK)}, \kappa_{bg} = 0.08 \:{(W/mK)}$ and $\kappa_f = 0.066 \:{(W/mK)}$. The mass densities are $\rho_s = 2650 \:{(kg/m^3)}, \rho_g = 1.836 \:{(kg/m^3)}$ and $\rho_f = 1000 \:{(kg/m^3)}$. The simulation has a total time of 80 s with increments of $\Delta t=$0.1 s. Figure \ref{fig:temperature_plots} shows the temperature evolution for the three phases at three different times. The length scale for fluid thermal conductivity $l_g = 1$ mm and variation of pore size is $\omega =2 \: \text{mm} - 1.95(y/th) \: \text{mm}$ where $th=6$ mm is the thickness of the domain. Figure \ref{fig:temperature_2} shows the temperature along the thickness of the domain at the center ($x = 0.5L$).

\clearpage
%========================================================================
% Bibliography
%========================================================================
%\clearpage
\bibliographystyle{abbrv} % Here the bibliography 		  %
\bibliography{main}

@article{YANG20104316,
title = {Analysis of temperature gradient bifurcation in porous media – An exact solution},
journal = {International Journal of Heat and Mass Transfer},
volume = {53},
number = {19},
pages = {4316-4325},
year = {2010},
issn = {0017-9310},
doi = {https://doi.org/10.1016/j.ijheatmasstransfer.2010.05.060},
url = {https://www.sciencedirect.com/science/article/pii/S0017931010003029},
author = {Kun Yang and Kambiz Vafai}
}

@article{AMIRI1994939,
title = {Analysis of dispersion effects and non-thermal equilibrium, non-Darcian, variable porosity incompressible flow through porous media},
journal = {International Journal of Heat and Mass Transfer},
volume = {37},
number = {6},
pages = {939-954},
year = {1994},
issn = {0017-9310},
doi = {https://doi.org/10.1016/0017-9310(94)90219-4},
url = {https://www.sciencedirect.com/science/article/pii/0017931094902194},
author = {A. Amiri and K. Vafai}
}

@article{alnaes2015fenics,
  title={The FEniCS project version 1.5},
  author={Aln{\ae}s, Martin and Blechta, Jan and Hake, Johan and Johansson, August and Kehlet, Benjamin and Logg, Anders and Richardson, Chris and Ring, Johannes and Rognes, Marie E and Wells, Garth N},
  journal={Archive of numerical software},
  volume={3},
  number={100},
  year={2015}
}

@book{phillips2005finite,
  title={Finite element methods in linear poroelasticity: Theoretical and computational results},
  author={Phillips, Phillip Joseph},
  year={2005},
  publisher={The University of Texas at Austin}
}

@Article{polym15030728,
AUTHOR = {Dei Sommi, Andrea and Lionetto, Francesca and Maffezzoli, Alfonso},
TITLE = {An Overview of the Measurement of Permeability of Composite Reinforcements},
JOURNAL = {Polymers},
VOLUME = {15},
YEAR = {2023},
NUMBER = {3},
ARTICLE-NUMBER = {728},
URL = {https://www.mdpi.com/2073-4360/15/3/728},
PubMedID = {36772028},
ISSN = {2073-4360},
DOI = {10.3390/polym15030728}
}

@article{NAIK201422,
title = {Permeability characterization of polymer matrix composites by RTM/VARTM},
journal = {Progress in Aerospace Sciences},
volume = {65},
pages = {22-40},
year = {2014},
issn = {0376-0421},
doi = {https://doi.org/10.1016/j.paerosci.2013.09.002},
url = {https://www.sciencedirect.com/science/article/pii/S0376042113000869},
author = {N.K. Naik and M. Sirisha and A. Inani}
}

@article{majumdar1993,
    author = {Majumdar, A.},
    title = {Microscale Heat Conduction in Dielectric Thin Films},
    journal = {Journal of Heat Transfer},
    volume = {115},
    number = {1},
    pages = {7-16},
    year = {1993},
    month = {02},
    issn = {0022-1481},
    doi = {10.1115/1.2910673}
}

@article{NAIRN199763,
title = {On the use of shear-lag methods for analysis of stress transfer in unidirectional composites},
journal = {Mechanics of Materials},
volume = {26},
number = {2},
pages = {63-80},
year = {1997},
issn = {0167-6636},
doi = {https://doi.org/10.1016/S0167-6636(97)00023-9},
author = {John A. Nairn}
}

@article{xia2002shear,
  title={Shear-lag versus finite element models for stress transfer in fiber-reinforced composites},
  author={Xia, Z and Okabe, T and Curtin, WA},
  journal={Composites Science and Technology},
  volume={62},
  number={9},
  pages={1141--1149},
  year={2002},
  publisher={Elsevier}
}

@article{beyerlein1996comparison,
  title={Comparison of shear-lag theory and continuum fracture mechanics for modeling fiber and matrix stresses in an elastic cracked composite lamina},
  author={Beyerlein, Irene J and Phoenix, S Leigh and Sastry, Ann Marie},
  journal={International Journal of Solids and Structures},
  volume={33},
  number={18},
  pages={2543--2574},
  year={1996},
  publisher={Elsevier}
}

@article{BOWEN19801129,
title = {Incompressible porous media models by use of the theory of mixtures},
journal = {International Journal of Engineering Science},
volume = {18},
number = {9},
pages = {1129-1148},
year = {1980},
issn = {0020-7225},
doi = {https://doi.org/10.1016/0020-7225(80)90114-7},
url = {https://www.sciencedirect.com/science/article/pii/0020722580901147},
author = {Ray M. Bowen}
}

@article{ERINGEN1965197,
title = {A Continuum theory of chemically reacting media—I},
journal = {International Journal of Engineering Science},
volume = {3},
number = {2},
pages = {197-212},
year = {1965},
issn = {0020-7225},
doi = {https://doi.org/10.1016/0020-7225(65)90044-3},
url = {https://www.sciencedirect.com/science/article/pii/0020722565900443},
author = {A.Cemal Eringen and John D. Ingram}
}
\index{Bibliography@\emph{Bibliography}}%
\end{document}